\documentclass{article}

\def\giorno{8 November 2026}

\usepackage{color}
\usepackage{amsmath,amsfonts,amssymb}
\usepackage{graphicx}
\usepackage{cite}
\usepackage{booktabs}
\def\a{\alpha}
\def\b{\beta}
\def\ga{\gamma}

\def\s{\sigma}

\def\om{\omega}

\def\H{{\mathcal H}}

\def\R{{\bf R}}

\def\T{{\rm T}}

\def\La{\Lambda}

\def\pa{\partial}

\def\<{\langle}
\def\>{\rangle}

\def\=#1{\bar #1}
\def\~#1{\widetilde #1}

\def\.#1{\dot #1}
\def\^#1{\widehat #1}

\def\"#1{\ddot #1}

\def\d{{\mathtt d}}
\def\interno{\cdot}
\def\w{\wedge}

\def\bar#1{\overline{#1}}

\def\beq{\begin{equation}}
\def\eeq{\end{equation}}

\def\beql#1{\begin{equation}\label{#1}}

\def\EOR{\hfill $\odot$}

\def\eqref#1{(\ref{#1})}

\def\salta#1{ }

\def\ri{i}
\def\wdh{\widehat}
\numberwithin{equation}{section}

\begin{document}

\title{A hyperhamiltonian approach to symmetry of the Dirac equation}

\author{G. Gaeta$^{1,2}$ \ and \ M.A. Rodr\'iguez$^3$\\
{} \\
${}^1$ {\it Dipartimento di Matematica, Universit\`a degli Studi di Milano,} \\ {\it via Saldini 50, 20133 Milano (Italy)} \\ and {\it INFN Sezione di Milano}  \\
${}^2$ {\it SMRI, 00058 Santa Marinella (Italy)} 
\\ {\tt giuseppe.gaeta@unimi.it} \\
{ } \\
${}^3$ {\it Departamento de F\'isica Te\'orica}\\ {\it Universidad Complutense de Madrid,} \\
{\it Plaza de las Ciencias 1,  28040 Madrid (Spain)} \\
{\tt rodrigue@ucm.es} 
}

\date{\giorno}

\maketitle

\begin{abstract}

\noindent The Dirac equation is the simplest wave equation invariant under the Poincar\'e group, and was indeed determined by Dirac precisely by requiring this invariance. Moreover it is invariant under dilations of the wave function, as follows by its linear character. It is known that the Dirac equation possesses, besides these, other symmetries, which in this sense are the special part of the Dirac equation's full symmetry. In previous work we have shown that the Dirac equation can be cast in hyperhamiltonian terms. Here we show that the symmetries can also be described in hyperhamiltonian terms.

\noindent

\end{abstract}

\vfill

\newpage

\tableofcontents

\newpage


\section{Introduction} 

The linear Dirac equation for (free) fermions 
\beql{eq:dirac} 
 (\ri \, \hbar \, \gamma^\mu \, \pa_\mu \ - \ m \, c   ) \ \psi \ = \ 0  
\eeq
is one of the most fundamental equations of Physics \cite{Dirac,Landau,BD,Messiah,Zee,Thal}. 
In the massless case, this is also known as the Majorana-Weyl equation, 
\beql{eq:MW}
\ri \, \hbar \, \gamma^\mu \, \pa_\mu  \, \psi \ = \ 0 \ . 
\eeq

Here $\psi = \psi (x)$ is the spinor wave function (four complex components) for a spin 1/2 particle of mass $m$, the constants $i,c,\hbar$ have their usual meaning, $\pa_\mu$ is the derivative w.r.t.~the coordinate $x^\mu$ with $\mu = 0,1,2,3$ and $x^0 =c\, t$, $\gamma^\mu$ are the Dirac gamma matrices, with $\gamma^5 =\ri\, \gamma^0 \gamma^1 \gamma^2 \gamma^3$, and sum over repeated indices is understood. It may be noted that a different notation is also in use in the literature, i.e. these equations are also written in terms of $\overline{\psi} := \psi^+ \gamma^0$. 

In view of the relevance of these equations for Physics, it is no surprise that their invariance properties have been extensively studied; see e.g. \cite{Bar,PuPl,KMW,Ibr69,FZ152,FZ155,FZbook,FNbook}. 
Some results about this matter will be recalled in Section \ref{sec:symmetry} below.

In a previous paper  \cite{GRDirac} we have provided a hyperhamiltonian description of the Dirac equation (and of other equations related to magnetic systems). Here, ``hyperhamiltonian'' refers to a generalization of Hamiltonian Mechanics, in which the role of the symplectic form is taken by an ordered triple of symplectic forms satisfying certain relations between themselves, so that they define a \emph{hyperKahler structure} \cite{AtHit,Hit} (sometimes -- albeit not with full precision -- also referred to as \emph{quaternionic structure}, or \emph{hypercomplex structure}; see e.g. \cite{QSMP2} for details, also about subtle differences between these notions). We refer to the literature \cite{GM,MT,GR0,GR1,GR2,GR3,GR4,GR5,GRbook} for details on this, while some basic facts and formulas relevant for our present discussion will be recalled in Section \ref{sec:hyperhamiltonian} below. 

The goal of this note is to discuss if, and to which extent, the vector fields spanning the symmetry algebra of the Dirac equations can themselves be cast in hyperhamiltonian form. It is an inherent limitation of our hyperhamiltonian approach that we will be only able to deal with \emph{local} Lie symmetries; that is, we will not consider \emph{nonlocal} Lie symmetries, nor non-Lie symmetries (which exist, and are discussed e.g. in \cite{FZbook}).

It should also be noted that among the symmetries of the Dirac equation, some are ``obvious'': the Dirac equation is built to display relativistic invariance (so the Poincar\'e group is obviously a symmetry group for it), and also the scale symmetry -- and the symmetry embodied in the superposition principle -- related to linearity is obviously present. Note however that the scale symmetry (by this we mean scaling of the wave function) disappear if we require the wave functions to be normalized. We also note in passing that the Lorentz transformations can be described in quaternionic terms, as already known by Dirac \cite{DL}. 

Somewhat surprisingly, these physically necessary symmetries do not exhaust the Dirac equation symmetry algebra, which also contains different, special, symmetry generators. We will mainly focus on these other special symmetries. We will show that these can be cast in hyperhamiltonian form.


\subsection{Notation}

As stated above, our discussion will be set in four-dimensional space with coordinates $x^\mu$, where $\mu = 0,\ldots,3$ and $x^0 =c\, t$; the Minkowski metric will be $(+1,-1,-1,-1)$. The (spinorial) wave function $\psi$ will have components $\psi^a$, which for spin $s=1/2$ entails $a=1,\ldots,4$.

In denoting partial derivatives we will adopt a simplified notation, with 
\beq
\pa_\mu \ = \ \frac{\pa}{\pa x^\mu} \ , \ \ \ \pa_a \ = \ \frac{\pa}{\pa \psi^a} \ , \ \ \ \pa^*_a \ = \ \frac{\pa}{\pa (\psi^*)^a} \ . 
\eeq

We use the standard form of Pauli matrices \cite{Dirac,Landau,Messiah,BD,Zee}, i.e.
\beql{eq:Pauli} 
\s^1 \ = \ \begin{pmatrix} 0 & 1 \\ 1 & 0 \end{pmatrix} \ , \ \ 
\s^2 \ = \ \begin{pmatrix} 0 & -\ri \\\ri & 0 \end{pmatrix} \ , \ \ 
\s^3 \ = \ \begin{pmatrix} 1 & 0 \\ 0 & - 1 \end{pmatrix} \ .  \eeq
The gamma matrices are then given, in the standard (Dirac) representation (we will also consider the Majorana and spinor representations, see below) by
\beql{eq:gammastandard} \gamma^0 \ = \ \begin{pmatrix} \sigma^0 & 0 \\ 0 & - \sigma^0 \end{pmatrix} \ , \ \ 
\gamma^k \ = \ \begin{pmatrix} 0 & \s^k \\ - \s^k & 0 \end{pmatrix}, \quad  k=1,2,3 \ ,  \eeq
where $\sigma^0$ is the two-dimensional identity matrix. This representation entails
\beql{eq:gamma5} \gamma^5 \ = \ \begin{pmatrix} 0 & \sigma^0 \\ \sigma^0 & 0 \end{pmatrix} \ . \eeq
We will also write, for later reference, 
\beql{eq:Smunu} S_{\mu \nu} \ := \ \frac14 \ \left( \ga_\mu \, \ga_\nu \ - \ \ga_\nu \, \ga_\mu \right)\ , \quad  
S_{\nu \mu} \ = \ - \ S_{\mu \nu} \ . 
\eeq

As already mentioned, the notation
 \beql{eq:psibar} \overline{\psi} \ := \  \psi^+ \, \gamma^0 \eeq is often used in the literature; we will generally use the explicit form for this.

In the hyperhamiltonian formalism, it is convenient to work in real spaces rather than in complex ones (we will however keep to the complex notation occasionally, to avoid having to display exceedingly large -- that is, eight-dimensional -- matrices). With the standard isomorphism between ${\bf C}^n$ and ${\bf R}^{2n}$, in which
\beql{eq:i:J}\ri \ \simeq \ J \ = \ \begin{pmatrix} 0 & - 1 \\ 1 & 0 \end{pmatrix} \ , \eeq
we then write 
\beq \label{eq:realsigma}
\Sigma^1 \ = \ \begin{pmatrix} 0 & \sigma^0 \\ \sigma^0 & 0   \end{pmatrix}  ,  \quad 
\Sigma^2 \ = \ \begin{pmatrix} 0 & -J \\
J & 0 \end{pmatrix} , \quad 
\Sigma^3 \ = \ \begin{pmatrix} \sigma^0 & 0   \\ 0 & -\sigma^0\end{pmatrix}   \eeq
for the Pauli matrices. The Dirac matrices become then eight-dimensional real matrices (which we do not write explicitly); their form is still given by \eqref{eq:gammastandard}, \eqref{eq:gamma5}, where now $\sigma^0$ should be understood as the four-dimensional identity matrix. 

Note that the standard commutation relations for the Pauli matrices:
$$[\sigma^j,\sigma^k]=2\, \ri \, \epsilon_{jkl}\, \sigma^k $$ are substituted, for the real forms in \eqref{eq:realsigma}, by
$$[\Sigma^j,\Sigma^k]=2\,C\,\epsilon_{jkl}\,\Sigma^l,\quad C= \begin{pmatrix} J& 0 \\ 0 & J\end{pmatrix} \ . $$

In the following we will not deal explicitly with the well known Poincar\'e and the conformal groups in four dimensional Minkowski space; we will however need to refer to these, which will be denoted as $P(1,3)$ and $C(1,3)$ respectively.


\subsection{Acknowledgements}

GG acknowledges the continuing support of INFN (Project MMNLP -- Mathematical Methods of Nonlinear Physics) and of GNFM-INdAM; he also thanks SMRI for hospitality. MAR acknowledges the support of MICIU (Spain) and FEDER (UE), project PID2024-156610NB-I00.



\section{Local Lie symmetries of the Dirac equation}
\label{sec:symmetry}

We will now briefly review symmetries of the Dirac equations. As mentioned in the Introduction, we will focus on \emph{local Lie symmetries}  \cite{Olver,Olver2,Stephani,Bluman,CicGaeBook}. 

The Dirac equation is, by construction, a relativistic equation; in symmetry terms, this means it is covariant under the Poincar\'e group. This includes spatial rotations as well as space-time Lorenz (hyperbolic) rotations.

The nontrivial question is if it admits other symmetry generators besides those of the Poincar\'e group.

We quote here, with slight notational changes, two theorems by Fushchich and Zhdanov summarizing the result on this matter in the massive ($m \not= 0$) and massless ($m=0$) cases respectively. Theorems A and B are respectively theorems 1.1.1 and 1.1.2 of \cite{FZbook}; the computations constituting a proof of these can be found e.g. in \cite{FZbook,Ibr69}. 

We will give a  detailed computation with different representations of Dirac $\gamma$ matrices  for the wave functions vertical symmetries  in the Appendix \ref{sec:symDirac2}, for completeness and for convenience of the reader.

\medskip\noindent
{\bf Theorem A.} {\it The maximal local invariance group of the Dirac equation \eqref{eq:dirac} is the 14-parameter group 
\beq 
G_1 \ = \ P(1,3) \otimes V(4) \ , 
\eeq 
where $P(1,3)$ is the Poincar\'e group, and $V(4)$ is the 4-parameter group of transformations in the space $(\psi^*,\psi)$ generated by the operators (in the standard (Dirac) representation of the $\gamma$ matrices): 
\beql{eq:genV4} 
\begin{aligned}
Q_0 &= \psi^a \, \pa_a \ + \ (\psi^*)^a \, \pa_a^*  \ , \\
Q_1 &=\ri \, \left[ \psi^a \, \pa_a \ - \ (\psi^*)^a \, \pa_a^* \right] \ , \\
Q_2 &= (\ga^2 \psi^* )^a \, \pa_a \ - \ (\ga^2 \psi)^a \, \pa_a^* \ , \\
Q_3 &=\ri \, \left[ (\ga^2 \psi^* )^a \, \pa_a \ + \ (\ga^2 \psi)^a \, \pa_a^* \right] \ .  
\end{aligned} 
\eeq 
}

\medskip\noindent
{\bf Theorem B.} {\it The maximal local invariance group of the massless Dirac equation \eqref{eq:MW} is the 23-parameter group 
\beq 
G_2 \ = \ C(1,3) \otimes V(8)\ , 
\eeq 
where $C(1,3)$ is the conformal group,  and $V(8)$ is the 8-parameter group of transformations in the space $(\psi^*,\psi)$ generated by the operators $\{ Q_0,\ldots,Q_3\}$ given above and moreover 
\beql{eq:genV8} 
\begin{aligned}
Q_4 &= (\ga^5 \psi )^a \, \pa_a \ + \ (\ga^5 \psi^*)^a \, \pa_a^* \ , \\
Q_5 &=\ri \, \left[ (\ga^5 \psi )^a \, \pa_a \ - (\ga^5 \psi^*)^a \, \pa_a^* \right] \ , \\
Q_6 &= (\ga^2 \ga^5 \psi^* )^a \, \pa_a \ - \ (\ga^2 \ga^5 \psi)^a \, \pa_a^* \ , \\
Q_7 &=\ri \, \left[ (\ga^2 \ga^5 \psi^* )^a \, \pa_a \ +   \ (\ga^2 \ga^5 \psi)^a \, \pa_a^* \right] \ . 
\end{aligned} \eeq
}

It is worth mentioning here that the above symmetries involve the spinor $\psi$ and its complex conjugate $\psi^*$. In fact, an analysis of the symmetries of the equation \eqref{eq:dirac} (vertical symmetries, leaving unchanged the coordinates) shows that there is no symmetry, except the trivial one, the identity. 

We note that if we make a linear transformation of the wavefunction $\psi$:
$\psi' =A \psi$, the equation becomes:
$$
(\ri\,\hbar\,\gamma^{\mu}\,\pa_{\mu}-m\,c\,I_4)\,A^{-1}\psi'=0
,\quad (i\,\hbar\,A\,\gamma^{\mu}\,A^{-1}\,\pa_{\mu}-m\,c\,I_4)\,\psi'=0\ .
$$
In order to leave the equation invariant, the matrix $A$ should satisfy (in the massive case) $A\,\gamma^{\mu}\,A^{-1}=\gamma^{\mu}$, $\mu=0,1,2,3$. We look for $4\times 4$ matrices commuting with $\gamma$ matrices:
$[A,\gamma^{\mu}]=0$, $\mu=0,1,2,3$. The $\gamma$ matrices generate an irreducible representation of the Clifford algebra and then, by Schur's  lemma,  the only matrices commuting with all of them are the multiples of the identity matrix: $A=\lambda\,I_4$.

\medskip\noindent{\bf Remark 1.} As well known, and already mentioned, Poincar\'e invariance is built-in in the massive Dirac equation (this was actually determined precisely requiring Poincar\'e invariance) and, together with conformal invariance, in the massless one; thus the ``interesting'' parts in the symmetry algebra determined by the two Theorems are precisely the $V(4)$ and $V(8)$ factors. We will hence focus on these, disregarding the $P(1,3)$ and $C(1,3)$ factors. It may be worth remarking explicitly also that $Q_0$ generates the dilation group, and is related to the linear nature of the equations \eqref{eq:dirac} and \eqref{eq:MW}. As already mentioned, $Q_0$  should be discarded if we work with \emph{normalized} wave functions. \EOR

 \medskip\noindent{\bf Remark 2.} The precise structure of the Lie algebras generated by the vector fields \eqref{eq:genV4} and \eqref{eq:genV8} will be discussed in Appendix \ref{sec:symDirac2}.\EOR


\section{Symmetry of the Dirac equation in the momentum representation\label{sec:symDirac1}}

Since our purpose is to discuss symmetry transformations of the Dirac equation in hyperhamiltonian form, following \cite{GRDirac} we will first write equation \eqref{eq:dirac} as:
\beq 
\ri\,\hbar\,\frac{\pa \psi }{\pa x^0}=\left(-\ri\,\hbar\, \gamma^0\, \gamma^j\, \pa_j+m\,c\, \gamma^0\right)\,\psi=\left(-\ri\,\hbar\, \alpha^j\, \pa_j+m\,c\, \beta\right)\,\psi\ ,
\eeq
where $\vec{\alpha}=(\gamma^0)^{-1} \vec{\gamma}=\gamma^0\,\vec{\gamma}$, $\beta = (\gamma^0)^{-1}=\gamma^0$,  and then we take the Fourier transform in the spatial variables, i.e.  
$$ \pa_j \to \ri\,p_j/\hbar \ , \ \ \psi(t,\vec{x}\,)\to \wdh{\psi}(t,\vec{p}\,)\ . $$
We will use $\wdh{\psi}$ for the wavefunction in this section to remark we are working with the Fourier transform.

This Fourier transform yields
\beq 
\ri\,\hbar\,\frac{\pa\wdh{\psi}}{\pa x^0}=\left( \vec{\alpha}\cdot \vec{p}+m\,c\, \beta\right)\wdh{\psi}\ .
\eeq

We now write the above equation in a real form using the pure imaginary matrices of the Majorana representation ($\alpha_k^*=\alpha_k$, $\beta^*=-\beta$):
\beq 
\begin{aligned}
&\gamma_M^0=\begin{pmatrix}0 & \sigma^2\\ \sigma^2 & 0\end{pmatrix},&&
\gamma_M^1=\begin{pmatrix}\ri\,\sigma^3 & 0\\ 0 &\ri\,\sigma^3\end{pmatrix}, \\
&\gamma_M^2=\begin{pmatrix}0 & -\sigma^2 \\ \sigma^2& 0\end{pmatrix},&&
\gamma_M^3=\begin{pmatrix}-\ri\, \sigma^1 & 0\\ 0 & -\ri\,\sigma^1  \end{pmatrix},
\end{aligned}
\eeq
and:
\beq 
\begin{aligned}
&\beta_M=\begin{pmatrix}0 & \sigma^2\\ \sigma^2 & 0\end{pmatrix},&&
\alpha_M^1=\begin{pmatrix} 0 & -\sigma^1\\ -\sigma^1 & 0\end{pmatrix},\\
&\alpha_M^2=\begin{pmatrix}\sigma^0 &  0 \\ 0 & -\sigma^0\end{pmatrix},&&
\alpha_M^3=\begin{pmatrix}0 & -\sigma^3\\ - \sigma^3 & 0  \end{pmatrix}.
\end{aligned}
\eeq

 The equations for $\wdh{\psi}$ and $\wdh{\psi}^*$ are:
\beql{eq:Fourier}
\begin{cases}
\dfrac{\ri\,\hbar}{c}\,\dfrac{\pa\wdh{\psi}}{\pa t}=\left( \vec{\alpha}_M\cdot \vec{p}+m\,c\, \beta_M\right)\wdh{\psi}\ ,  
\\[10pt]
\dfrac{\ri\,\hbar}{c}\,\dfrac{\pa\wdh{\psi}^*}{\pa t}=\left(- \vec{\alpha}_M\cdot \vec{p}+m\,c\, \beta_M\right)\wdh{\psi}^*\ ,
\end{cases}
\eeq
and if we define
\beq 
\wdh{\psi}_{a}= \wdh{\xi}_{a}+\ri\, \wdh{\zeta}_{a},\quad a=1,\ldots ,4\ ,
\eeq
\beq 
\wdh{\Phi}=\left( 
\wdh{\xi}_1,\, \wdh{\zeta}_1,\;\; 
\wdh{\xi}_2,\, \wdh{\zeta}_2,\;\; 
\wdh{\xi}_3,\, \wdh{\zeta}_3,\;\; 
\wdh{\xi}_4,\, \wdh{\zeta}_4
\right)^T\ ,
\eeq
the equation becomes
\begin{equation}\label{eq:FourierHam}
\frac{\hbar}{c}\frac{\pa \wdh{\Phi}}{\pa  t}=H_M\, \wdh{\Phi},
\end{equation}
where
\beq\label{ec:ham1}
H_M=\sum_{i=1}^3\Pi_M^i\,p_i + \Pi_M^0 
 \,m\, c
\eeq
and\footnote{We will use the tensor product notation to simplify the writing of these 8-dimensional matrices. The tensor product in this case is the standard Kronecker product of matrices.}
\beq 
\begin{aligned}
&\Pi_M^0= \sigma^1\otimes (\gamma_M^0\gamma_M^5) ,&&\Pi_M^1= -\ri\,\sigma^1\otimes \gamma_M^0 ,\\
&\Pi_M^2=\sigma^3\otimes (\gamma_M^1\gamma_M^3) ,&&\Pi_M^3=-\ri\,\sigma^1\otimes \gamma_M^5\ . 
\end{aligned}
\eeq
  
In the standard representation of the $\gamma$ matrices, the Hamiltonian \eqref{ec:ham1} is written in the same form, with $\Pi^{\mu}$ matrices:
\beq 
\begin{aligned}
&  \Pi_D^0=  \sigma^3\otimes (\gamma_D^1\gamma_D^3), &&\Pi_D^1= \ri\,\sigma^1\otimes (\gamma_D^0\gamma_D^2) ,\\ 
&\Pi_D^2= \sigma^1\otimes (\gamma^5_D \gamma_D^0)  ,   &&\Pi_D^3= \ri\,\sigma^1\otimes (\gamma_D^2\gamma_D^5) \  .
\end{aligned}
\eeq

Finally, in the spinor representation, where
\beq 
\ga^0 \ = \ \begin{pmatrix} 0 &\sigma^0 \\ \sigma^0 & 0 \end{pmatrix} \ , \ \ \ \gamma^k \ = \ \begin{pmatrix} 0& -\sigma^k \\ \sigma^k & 0 \end{pmatrix} \ . 
\eeq 
the matrices $\Pi_S^{\mu}$ are:
\beq 
\begin{aligned}
&  \Pi_S^0=  \sigma^1\otimes (\gamma_S^1\gamma_S^3), &&\Pi_S^1= \ri\,\sigma^3\otimes (\gamma_S^2\gamma_S^5) ,\\ 
&\Pi_S^2= \sigma^3\otimes (\gamma^0_S \gamma_S^5)  ,   &&\Pi_S^3= \ri\,\sigma^3\otimes (\gamma_S^0\gamma_S^2) \  .
\end{aligned}
\eeq


\subsection{Symmetries}

Our aim is to discuss linear symmetries (in the wave function) of the Dirac equation in the formulation given above \eqref{eq:FourierHam} ($\hbar=c=1$):
\beq
\pa_t\wdh{\Phi} =H\,\wdh{\Phi} \ .
\eeq
Then, if
\beq
\wdh{\Phi} '=\Lambda\,\wdh{\Phi}\  ,
\eeq
where $\Lambda$ is a real regular $8\times 8$ matrix, we get:
\beq
\pa_t\,\Lambda^{-1}\,\wdh{\Phi}'=H\,\Lambda^{-1}\,\wdh{\Phi} ',\quad \pa_t\,\wdh{\Phi}'=\Lambda\, H\,\Lambda^{-1}\,\wdh{\Phi}'\ ,
\eeq
and then, $\Lambda$ should commute with $H$:
\beq
 [H,\Lambda]=0\ .
\eeq

\subsubsection{Massive case}
In the massive case  the commutation relations to be satisfied are:
\beq
 [\Pi_{\mu},\Lambda]=0,\quad \mu=0,1,2,3\ .
\eeq

The explicit form of the matrix $\Lambda$ can be easily found from the above equations. It depends on four real parameters which can be chosen to yield the following four matrices, in the Majorana representation:
\beql{ec:Majoranamassiverep}
\begin{aligned}
&L^M_0=\frac12\,\sigma^0\otimes I_4\equiv I_8,\quad 
&&L^M_1= \frac12\,\sigma^1\otimes (\gamma_M^1\,\gamma_M^2), \\
&L^M_2 =\frac12\,\sigma^1\otimes  (\gamma_M^2\,\gamma_M^3) ,\quad 
&&L^M_3 = \frac12\,\sigma^0\otimes ( \gamma_M^3\,\gamma_M^1 )\ .  
\end{aligned}
\eeq
and, in the standard representation: 
\beql{ec:Diracmassiverep}
\begin{aligned}
&L^D_0=\frac12\,\sigma^0\otimes I_4\equiv I_8,\quad 
&&L^D_1=-\frac12\,\sigma^1\otimes  \gamma_D^1 ,\\
&L^D_2=\frac12\,\sigma^1\otimes \gamma_D^3,\quad  
&&L^D_3=\frac12\,\sigma^0\otimes (\gamma_D^3\,\gamma_D^1)\ . 
\end{aligned}
\eeq
Similar expressions can be obtained for the spinor representation. The corresponding symmetry vector fields can be easily obtained from the matrices $L_i$ in the standard or Majorana representations. It can be shown that the matrix $L_0^D$ commutes with the other three  $L^D_i$, $i=1,2,3$ and that these latter matrices  (in any representation of the $\gamma$ matrices) are a basis of the Lie algebra $\mathrm{so}(3)$ (see Table \ref{table:tab1}). This is in contrast with the result in   Appendix \ref{sec:symDirac2} where the corresponding matrices generate the Lie algebra  $\mathrm{so}(2,1)$. The reason for this discrepancy lies in the use of the Fourier transform, which introduces a  factor $i$ in the structure of the matrices (see also Appendix \ref{sec:Fourier}).

\subsubsection{Massless case}
Consider the Dirac equation for $m=0$:
\beq\frac{\hbar}{c}\frac{\pa \wdh{\Phi}}{\pa t}=H\wdh{\Phi}
,\quad H= \sum_{i=1}^3\Pi_i\,p_i \ ,
\eeq
\beq
 [\Pi_{k},\Lambda]=0,\quad k=1,2,3\ .
\eeq
Imposing commutativity of the Hamiltonian with the symmetry matrices, we get, apart from the matrices given in \eqref{ec:Majoranamassiverep} and \eqref{ec:Diracmassiverep} for the massive case, the matrices in the Majorana representation
\beq
\begin{aligned}
&B_0^M=\frac{\ri}{2}\,\sigma^3\otimes \gamma_M^2,  
&&B_1^M=\frac12\,\sigma^2\otimes \gamma_M^1\ ,\\  
&B_2^M=\frac12\,\sigma^2\otimes \gamma_M^3,
&&B_3^M=\frac12\,\sigma^3\otimes (\gamma_M^5\,\gamma_M^0)\ . 
\end{aligned}
\eeq
and in the standard representation:
\beql{ec:Diracmasslessrep}
\begin{aligned}
&B_0^D=\frac12\,\sigma^1\otimes I_4,
&&B_1^D=-\frac12\,\sigma^0\otimes  \gamma_D^1\ ,\\
&B_2^D=\frac12\,\sigma^0\otimes  \gamma_D^3,   
&&B_3^D=-\frac12\,\sigma^1\otimes (\gamma_D^1\,\gamma_D^3)\ . 
\end{aligned}
\eeq

As in the massive case, the algebra is not the same we get for the massless Dirac equation in the configuration space (see Appendix \ref{sec:symDirac2}). As it is easy to prove, these matrices generate a Lie algebra of dimension 6 which is isomorhpic to the compact orthogonal algebra $\mathrm{so}(4)$ plus a two dimensional center. In fact, we can rewrite the $\mathrm{so}(4)$ basis as:
\beq
\begin{aligned}
&\kappa_1=\frac12(L^M_1+B^M_1),\quad
\kappa_2=\frac12(L^M_2-B^M_2),\quad
\kappa_3=\frac12(L^M_3+B^M_3)\\
&\rho_1=\frac12(L^M_1-B^M_1),\quad
\rho_2=\frac12(L^M_2+B^M_2),\quad
\rho_3=\frac12(L^M_3-B^M_3)\ .
\end{aligned}
\eeq
Then, $\{\kappa_1,\kappa_2,\kappa_3\}$ and $\{\rho_1,\rho_2,\rho_3\}$ generate two commuting $\mathrm{so}(3)$
algebras, showing the structure of $\mathrm{so}(4)\approx \mathrm{so}(3)\oplus \mathrm{so}(3)$.

\begin{table}
\begin{center}
\begin{tabular}{c|ccc}
      & $L_1$   & $L_2$  & $L_3$  \\
\midrule
$L_1$ & $0$     & $L_3$  & $-L_2$ \\
$L_2$ & $-L_3$  & $0$    & $L_1$  \\
$L_3$ & $L_2$   & $-L_1$ & $0$  
\end{tabular}
\caption{Commutation table of the symmetry matrices of the massive Dirac equation in the momentum configuration  (the Lie algebra $\mathrm{so}(3)$).\label{table:tab1}}
\end{center}\end{table}

\begin{table}
\begin{center}
\begin{tabular}{c|cccccccc}
        & $L_1$   & $L_2$  & $L_3$  & $B_1$   & $B_2$  & $B_3$\\
\midrule
$L_1$   & $0$     & $L_3$  & $-L_2$ & $0$     & $-B_3$ & $B_2$  \\
$L_2$   & $-L_3$  & $0$    & $L_1$  & $-B_3$  & $0$    & $B_1$  \\
$L_3$   & $L_2$   & $-L_1$ & $0$    & $-B_2$  & $B_1$  & $0$    \\
$B_1$   & $0$     & $B_3$  & $B_2$  & $0$     & $-L_3$ & $-L_2$ \\
$B_2$   & $B_3$   & $0$    & $-B_1$ & $L_3$   & $0$    & $-L_1$ \\
$B_3$   & $-B_2$  & $-B_1$ & $0$    & $L_2$   & $L_1$  & $0$ 
\end{tabular}
\caption{Commutation table of the symmetry matrices of the massless Dirac equation in the momentum configuration (the Lie algebra $\mathrm{so}(4)$).\label{table:tab2}}
\end{center}\end{table}


\section{Hyperhamiltonian dynamics}
\label{sec:hyperhamiltonian}

We will now briefly introduce the essential features of hyperhamiltonian dynamics, following \cite{GRDirac}; see \cite{GRbook} for further details. We will then discuss, again following \cite{GRDirac}, the description of the Dirac equation in such terms.


\subsection{General features}

In hyperhamiltonian dynamics, one considers first a hyperKahler manifold 
$$(M,g,J_\a),$$
i.e. a smooth $4n$-dimensional (real) manifold $M = M^{4 n}$ equipped with a hyperKahler structure; this means $M$ is equipped with a Riemannian metric $g$ and an ordered triple of complex structures $J_\a$ (these are (1,1) type antisymmetric tensor fields) satisfying the quaternionic relations 
\beql{eq:quat} J_\a \, J_\b \ = \ \epsilon_{\a \b \ga} \, J_\ga \ \ - \ \delta_{\a \b} \,I_4 \ . \eeq 
To each of these is associated a Kahler form $\om_\a$, satisfying 
\beql{eq:kahler} \om_\a (v,w) \ = \ g \left( v , J_\a w \right) \eeq
for any $v,w \in \T M$. These are closed forms,
\beq \d\, \om_\a \ = \ 0 \ ; \eeq
this also means they are symplectic forms, so that hyperKahler manfolds can also be seen as ``hypersymplectic manifolds''.

\medskip\noindent
{\bf Remark 3.} Two general classes of manifolds are immediately seen to admit hyper-Kahler structures: these are quaternionic manifolds ${\bf H}^n \simeq {\bf R}^{4 n}$; and the cotangent bundle $M = T_* V$  of any Hermitian manifold $V$. Other nontrivial hyperKahler manifolds are obtained from these via the hyperKahler quotient construction introduced by Hitchin {\it et al.} in \cite{HKLR}. \EOR
\bigskip

Consider now an ordered triple of arbitrary smooth functions $\H_\a : M \to \R$. We associate
with these a triple of vector fields by (no sum on $\a$)
\beq X_\a \interno \om_\a \ = \ \d \H_\a \ , \eeq
and define the hyperHamiltonian vector field $X$ on $M$ associated with the ordered triple $\{ \H_\a \}$ as the
sum of these,
\beq X \ = \ \sum_{\a=1}^3 \ X_\a \ . \eeq
It is trivial to check that the $X_\a$, and hence $X$, are uniquely defined.

It is worth  considering a local chart on $M$ and in there local coordinates $\{ x^1,\ldots, x^m \}$; we write as usual $\pa_i := (\pa / \pa x^i)$. The Riemannian metric $g$ will be represented in these coordinates by a (0,2) tensor field $g_{ij} (x)$, the complex structure $J_\a$ by a (1,1) tensor field $(Y_\a)_{ij} (x)$, and the symplectic form $\om_\a$ by a (0,2) tensor field $(K_\a)_{ij} (x)$, with (omitting from now on to explicitly indicate the $x$ dependence, for ease of notation) \beq \om_\a \ = \ \frac12 \ K_{ij} \ \d x^i \w \d x^j \ . \eeq 
The Kahler relation \eqref{eq:kahler} implies that
\beq (K_\a)_{ij} \  = \ g_{ik} \ (Y_\a)^k_{\ j} \ . \eeq
When expressed in local coordinates, the relation $X_\a \interno \om_\a = \d \H_\a$ (no sum on $\a$) yields
$$ { (K_\a^T)_{ij} \, X^j \ = \ \pa_i \H_\a } \ ; $$ recalling the $K_\a$ are non-degenerate (being associated to a symplectic form), we can introduce $\La_\a = ( K_\a^T )^{-1}$, and write 
$$ X_\a^i \ = \ (\La_\a)^{ij} \ \pa_j \H_\a $$ (no sum on $\a$). Summarizing, the hyperhamiltonian vector field is given in local coordinates by 
\beq \left( \sum_{\a=1}^3 \La_\a^{ij} \ \pa_j \H_\a \right) \ \pa_i \ . \eeq

A specially relevant case (also for our present application) is provided by Euclidean four-dimensional space, $M = {\bf R}^{4 n}$. In this case on each ${\bf R}^4$ building block there are \emph{two} standard hyperhamiltonian structures, corresponding to the two standard real irreducible representations of $\mathrm{su}(2)$ \cite{Kirillov}; we will denote the tensor fields associated to these as $\{ Y_\a \}$ and $\{ \^Y_\a \}$ respectively. They are given in explicit form by 
\begin{eqnarray*}  Y_1 &=& \begin{pmatrix} 0 & 1 & 0 & 0 \\-1 & 0 & 0 & 0 \\ 0 & 0 & 0 & 1 \\ 0 & 0 & -1 & 0 \end{pmatrix} \ , \ \ Y_2 \ = \ \begin{pmatrix} 0 & 0 & 0 & 1 \\ 0 & 0 & 1 & 0 \\ 0 & -1 & 0 & 0 \\ -1 & 0 & 0 & 0 \end{pmatrix} \ , \\ 
Y_3 &=& \begin{pmatrix} 0 & 0 & 1 & 0 \\ 0 & 0 & 0 & -1 \\ -1 & 0 & 0 & 0 \\ 0 & 1 & 0 & 0 \end{pmatrix} \ ;  \\
\^Y_1 &=& \begin{pmatrix} 0 & 0 & 1 & 0 \\ 0 & 0 & 0 & 1 \\ -1 & 0 & 0 & 0 \\ 0 & -1 & 0 & 0 \end{pmatrix} \ , \ \ 
\^Y_2 \ = \ \begin{pmatrix} 0 & 0 & 0 & -1 \\ 0 & 0 & 1 & 0 \\ 0 & -1 & 0 & 0 \\ 1 & 0 & 0 & 0 \end{pmatrix} \ , \\ 
\^Y_3 &=& \begin{pmatrix} 0 & -1 & 0 & 0 \\ 1 & 0 & 0 & 0 \\ 0 & 0 & 0 & 1 \\ 0 & 0 & -1 & 0 \end{pmatrix} \ .  \end{eqnarray*}
The corresponding symplectic forms are respectively
\begin{eqnarray*}
\om_1 &=& \d x^2 \w \d x^2 \ + \ \d x^3 \w \d x^4 \ , \ \
\om_2 \ = \ \d x^1 \w \d x^4 \ + \ \d x^2 \w \d x^3 \ , \\
\om_3 &=& \d x^1 \w \d x^3 \ + \ \d x^4 \w \d x^2 \ ; \\
\^\om_1 &=& \d x^1 \w \d x^3 \ + \ \d x^2 \w \d x^4 \ , \ \
\^\om_2 \ = \ \d x^4 \w \d x^1 \ + \ \d x^2 \w \d x^3 \ , \\
\^\om_3 &=& \d x^2 \w \d x^1 \ + \ \d x^3 \w \d x^4 \ .  \end{eqnarray*}

It is immediately checked that the $\{Y_\a \}$ and the $\{ \^Y_\a \}$ matrices realize the $\mathrm{su}(2)$ Lie algebra commutation relations, 
\beql{eq:su2cr} \left[ Y_\a , Y_\b \right] \ = \ \epsilon_{\a \b \ga} \, Y_\ga \ , \ \ \ \left[ \^Y_\a , \^Y_\b \right] \ = \ \epsilon_{\a \b \ga} \, \^Y_\ga \ ; \eeq on the other hand, these two $\mathrm{su}(2)$ representations do commute,
\beql{eq:su2schur} \left[ Y_\a , \^Y_\b \right] \ = \ 0 \ , \ \ \forall \a,\b \ . \eeq

\medskip\noindent
{\bf Remark 4.} Such a set of two dual hyperkahler structures in ${\bf R}^4$ was christened a ``Dirac structure'' in \cite{GRDirac}. \EOR

\medskip\noindent{\bf Remark 5.} 
It may be worth noting that the matrices $Y_\a$ and $\^Y_\a$ can be written (in a compact notation) in terms of the Pauli matrices and of the identity matrix $I$ as
\begin{eqnarray*}
Y_1 &=& \begin{pmatrix}\ri \s_2 & 0 \\ 0 &\ri \s_2 \end{pmatrix} \ , \ \ Y_2 \ = \ \begin{pmatrix} 0 & \s_1 \\ - \s_1 & 0 \end{pmatrix} \ , \ \ Y_3 \ = \ \begin{pmatrix} 0 & \s_3 \\ - \s_3 & 0 \end{pmatrix} \\
\^Y_1 &=& \begin{pmatrix} 0 &\ri \\ -\ri & 0  \end{pmatrix} \ , \ \ \^Y_2 \ = \ \begin{pmatrix} 0 & -\ri \s_2 \\ -\ri \s_2 & 0 \end{pmatrix} \ , \ \ \^Y_3 \ = \ \begin{pmatrix} -\ri \s_2 & 0 \\  0 &\ri \s_2 \end{pmatrix} \ . \end{eqnarray*}
Here $i$ should of course be understood as its two-dimensional real representative $J$, see \eqref{eq:i:J} above. \EOR


\subsection{The Walcher factorization principle}
\label{sec:WFP}

In the following we will deal with vector fields $X$ of the form 
$$ X \ = \ X_+ \ + \  X_- \ , \ \ \ [X_+ , X_- ] \ = 0 \ , $$ i.e. which are expressed as the sum of two commuting vector fields. 

Let us denote as $\varphi (X;x_0,t)$ the flow issued by $x_0$ under the vector field $X$, i.e. the solution to 
$$ \frac{dx}{dt} \ = \ X (x) \ , \ \ \ x(t_0 ) = x_0 \ . $$
Then the Walcher factorization principle \cite{WalFP} states that this can be expressed as 
\beq \varphi (X;x_0,t) \ = \ \varphi \left( X_+ ; \varphi(X_-,x_0,t),t \right) \ = \ \varphi \left( X_- ; \varphi(X_+,x_0,t),t \right) \ . \eeq
In other words, due to the commutation property, we can follow the flow under $X_\pm$ for a time $t$, and then the flow under $X_\mp$ again for a time $t$, and obtain the flow under $X$ at time $t$.

This factorization will be essential in our discussion of  the next  Section \ref{sec:lindirhh}.


\subsection{The Dirac equation in the hyperhamiltonian formalism}
\label{sec:lindirhh}

We will consider the Dirac equation in both the  standard  representation and in the Majorana one.

The wave function $\psi$ is a bispinor; it can thus be written as 
$$ \psi \ = \ \begin{pmatrix} \psi_+ \\ \psi_- \end{pmatrix} \ , $$ where each of $\psi_\pm$ is a two-component spinor, which can also be represented as a four dimensional real vector:
$$ \psi_\pm \ = \ \begin{pmatrix} \psi_\pm^1 \\ \psi_\pm^2 \end{pmatrix} \ \simeq \begin{pmatrix} \mathtt{Re} \psi_\pm^1 \\ \mathtt{Im} \psi_\pm^1 \\ \mathtt{Re} \psi_\pm^2 \\ \mathtt{Im} \psi_\pm^2 \end{pmatrix} \ . $$
This also entails that $\psi$ is represented as an eight-dimensional real vector.

Now we should decide which representation we are using for the Dirac equation; it is convenient to use the momentum representation. This means that momenta can to all effects be considered as constants (that is, $\psi$ will be the wave function for a spin 1/2 fermion with a \emph{given momentum}), and we set $\pi = (\pi_1 , \pi_2 , \pi_3)$. We will then write\footnote{We could of course interchange the two $\mathrm{su}(2)$ representations, i.e. swap ${\bf \^K}$ and ${\bf K}$ on the one hand, and (see later) $Y_1$ and $\^Y_1$ on the other hand.} 
\beq {\bf \^K} \ = \ \^Y_2 \, \pi_1 \ + \ \^Y_1 \, \pi_2 \ + \ \^Y_3 \, \pi_3 \ . \eeq

We can still use the standard representation of the $\ga$ matrices, see above, or the Majorana or spinor representations for these (see below).

\subsubsection{Standard representation}

Let us first consider the standard representation for the gamma matrices, see \eqref{eq:gammastandard}, \eqref{eq:gamma5} above. 
Then the Dirac equation in the presence of an electromagnetic potential $\Phi$ (the free Dirac equation is of course obtained setting $\Phi = 0$) reads
\beql{eq:diracstandhh} 
\hbar \ \pa_t \, \begin{pmatrix} \psi_+ \\ \psi_- \end{pmatrix} \ = \ \begin{pmatrix} (m\, c^2 \, + \, e\, \Phi ) \, Y_1 & - c \, {\bf \^K} \\- c \, {\bf \^K} & - \, (m \, c^2 \, + \, e\, \Phi ) \, Y_1 \end{pmatrix} \  \begin{pmatrix} \psi_+ \\ \psi_- \end{pmatrix} 
\eeq

Thus we have set the Dirac equation (in the standard momentum representation) in explicit hyperhamiltonian form. It should be stressed that strictly speaking the flow described by \eqref{eq:diracstandhh} is the sum of \emph{two} hyperhamiltonian flows, one with respect to the standard structure defined by the $\{ Y_\a \}$ (albeit only $Y_1$  is actually present), and one defined by the conjugate structure defined by the $\{\^Y_\a \}$. 

Given that the two irreps of $\mathrm{su}(2)$ actually commute, see \eqref{eq:su2schur} above, the Walcher factorization principle (see Section \ref{sec:WFP} above) can now be used to conclude that \emph{the flow of the Dirac equation \eqref{eq:diracstandhh} is the superposition of two hyperhamiltonian flows}.

\medskip\noindent{\bf Remark 6.} The fact that we need \emph{two} hyperhamiltonian structures should not be surprising: after all, a hyperhamiltonian structure is defined by three matrices, while the Dirac equation involves the four $\gamma$ matrices $\ga^\mu$. \EOR

\medskip\noindent{\bf Remark 7.} As we have noted before, the two conjugate $\mathrm{su}(2)$ representations commute with each other. This, together with the structure of the (quaternionic form of the) Dirac equation allows to state immediately that $Y_1$ will generate a symmetry for the equation. \EOR

\subsubsection{Spinor representation}

In the so called spinor representation, the $\gamma$ matrices are written as 
\beq 
\ga^0 \ = \ \begin{pmatrix} 0 & \sigma^0 \\ \sigma^0 & 0 \end{pmatrix} \ , \ \ \ \gamma^k \ = \ \begin{pmatrix}  0 & -\s^k \\ \s^k & 0 \end{pmatrix} \ . 
\eeq 
The Dirac equation reads now
\beql{eq:diracspinhh} 
\hbar \ \pa_t \, \begin{pmatrix} \psi_+ \\ \psi_- \end{pmatrix} \ = \ \begin{pmatrix} - c \, {\bf \^K} & (m\, c^2 \, + \, e\, \Phi ) \, Y_1 \\ (m\, c^2 \, + \, e\, \Phi ) \, Y_1 &  c \, {\bf \^K}  \end{pmatrix} \  \begin{pmatrix} \psi_+ \\ \psi_- \end{pmatrix} 
\eeq

This shows that the Dirac equation (in the spinor representation) is also set in explicit hyperhamiltonian form. 
The remarks presented above -- about two conjugate hyperhamiltonian structures appearing here, about the Dirac flow being the superposition of two commuting hyperhamiltonian flows, and about $Y_1$ being an obvious symmetry -- apply again.


\subsection{The massless Dirac equation in the hyperhamiltonian formalism}

In the case of zero electromagnetic potential, the $Y_1$ matrix is only coupled to the mass term in the Dirac equation, but this coupling is of course absent in the massless case. Thus for a free spin 1/2 particle of zero mass, the Dirac { (Majorana-Weyl)} equation is written in hyperhamiltonian terms by means of \emph{only one} of the two conjugate hyperhamiltonian structures, i.e. as 
\beql{eq:MWstandhh} \hbar \ \pa_t \, \begin{pmatrix} \psi_+ \\ \psi_- \end{pmatrix} \ = \ 
\begin{pmatrix} 0 & - c \, {\bf \^K} \\- c \, {\bf \^K} & 0  \end{pmatrix} \  \begin{pmatrix} \psi_+ \\ \psi_- \end{pmatrix} \eeq
in the standard representation, and as 
\beql{eq:MWspinhh} \hbar \ \pa_t \, \begin{pmatrix} \psi_+ \\ \psi_- \end{pmatrix} \ = \ \begin{pmatrix} - c \, {\bf \^K} & 0 \\ 0 &  c \, {\bf \^K}  \end{pmatrix} \  \begin{pmatrix} \psi_+ \\ \psi_- \end{pmatrix} \eeq
in the spinor representation. 

\medskip\noindent
{\bf Remark 8.} Note that in these cases, the full $\mathrm{su}(2)$ algebra generated by the $\{ Y_1 , Y_2 , Y_3 \}$ is a symmetry algebra for the equation. \EOR


\section{Symmetry of the Dirac equation in hyperhamiltonian formalism}
\label{sec:hhsymm}

We will now investigate if the Lie symmetry generators of the Dirac equation can be expressed as hyperhamiltonian vector fields. In this investigation, we will not consider the Poincar\'e group generators, which act on space-time variables, and only focus on the transformations acting in the bispinor space ${\bf C}^4 \simeq {\bf R}^8$. 

We also note that the linearity of the equation entails that it will be invariant under a scaling transformation $\psi \to \lambda \psi$, which surely cannot be given a hyperhamiltonian representation (as hyperhamiltonian vector fields are divergence free -- here we refer of course to the space of wavefunctions with coordinates $\psi^a$. As already noted, $Q_0$ should be disregarded if, as it usually happens, one wants to work with normalized wavefunctions.

In other words, our investigation focuses on the subgroups $V(4)$ and $V(8)$ given in Theorems A and B, and which constitute the substantial part of these Theorems. In the language of the geometrical approach to symmetries of differential equations \cite{Olver,Olver2,Bluman,Stephani,CicGaeBook}, we are focusing on \emph{vertical} symmetry vector fields.

We will now give an independent computation of the Lie symmetry algebra for the Dirac equation, based on our hyperhamiltonian representation of the latter. (A computation in the configuration space is provided, for the convenience of the reader, in the Appendix \ref{sec:symDirac2}.)


\subsection{Massive Dirac equation}

As we have seen above, see eq.\eqref{eq:diracstandhh}, in the standard representation the Dirac equation reads
\beq \hbar \, \pa_t\, \psi \ = \ L \, \psi \ , \eeq
where
\beq L \ = \ \begin{pmatrix} m \, c^2 \, Y_1 & - c\, {\bf \^K} \\- c\, {\bf \^K} & - m \, c^2 \, Y_1 \end{pmatrix} \ . \eeq

Lie-point symmetries of the equation will be given by vector fields 
\beql{eq:dirsymmM} V \ = \ (M \psi)^i \, \frac{\pa}{\pa \psi^i} \eeq
(where the sum is over $i=1,\ldots,8$) with $M$ a matrix commuting with $L$. 

In looking for symmetries of this we will assume $m \not= 0$ and deal with an arbitrary momentum $\pi$.

Let us consider general block matrices 
\beql{eq:Mam0} M \ = \ \begin{pmatrix} A & B \\ C & D \end{pmatrix} \ , \eeq and study the condition $[L,M] = 0$. By elementary algebra it is seen that (to hold with general $m$, $\pi$ and hence ${\bf \^K}$) ) this requires 
\begin{eqnarray}
& & [A,Y_1] \ = \ 0 , \ \{ B , Y_1 \} \ = \ 0 , \ \{ C , Y_1 \} \ = \ 0 , \ [D,Y_1] \ = \ 0 \ ; \label{eq:CR1}\\
& & A {\bf \^K} \ = \ {\bf \^K} D , \ D {\bf \^K} \ = \ {\bf \^K} A , \ B {\bf \^K} \ = \ {\bf \^K} C , \ C {\bf \^K} \ = \ {\bf \^K} B \ . \label{eq:CR2} \end{eqnarray}

The most general matrices $A,D$ satisfying \eqref{eq:CR1} are of the form
\beq A \ = \ \left(
\begin{array}{llll}
 a_{11} &
   a_{12} &
   a_{13} &
   a_{14} \\
 -a_{12} &
   a_{11} &
   -a_{14} &
   a_{13} \\
 a_{31} &
   a_{32} &
   a_{33} &
   a_{34} \\
 -a_{32} &
   a_{31} &
   -a_{34} &
   a_{33}
\end{array}
\right) , \quad D \ = \ \left(
\begin{array}{llll}
 d_{11} &
   d_{12} &
   d_{13} &
   d_{14} \\
 -d_{12} &
   d_{11} &
   -d_{14} &
   d_{13} \\
 d_{31} &
   d_{32} &
   d_{33} &
   d_{34} \\
 -d_{32} &
   d_{31} &
   -d_{34} &
   d_{33}
\end{array}
\right) \ . \eeq

Now we should take care of \eqref{eq:CR2}, or more precisely of the intertwining relations between $A$ and $D$ enforced by \eqref{eq:CR2}. As these must hold for any choice of ${\bf \^K}$, they must hold for each of the $\^Y_\a$. Actually, the intertwining relations (recalling $\^Y_\a^{-1} = - \^Y_\a$ can be recast in the form
$$ D \ = \ - \, \^Y_\a \, A \, \^Y_\a \ . $$ 
When we implement this requirement for $\a = 1,2,3$ (for $A$ and $D$ as given above) we arrive at
\beq A \ = \ D \ = \ \left(
\begin{array}{llll}
 k_0 & - k_1 & 0 & 0
   \\
 k_1 & k_0 & 0 &
   0 \\
 0 & 0 & k_0 & - k_1
   \\
 0 & 0 & k_1 &
   k_0
\end{array}
\right)  \eeq
with $k_0,k_1$ arbitrary constants. This relation can be rewritten as
\beq A \ = \ D \ = \ k_0 \,I \ + \ k_1 \, \^Y_2 \ . \eeq

Let us now consider the matrices $B$ and $C$, which are required by \eqref{eq:CR1} to anticommute with $Y_1$. With standard algebra we find that the most general real matrices satisfying this requirement are
\beq B \ = \ \left(
\begin{array}{llll}
 b_{11} &
   b_{12} &
   b_{13} &
   b_{14} \\
 b_{12} &
   -b_{11} &
   b_{14} &
   -b_{13} \\
 b_{31} &
   b_{32} &
   b_{33} &
   b_{34} \\
 b_{32} &
   -b_{31} &
   b_{34} &
   -b_{33}
\end{array}
\right) \ , \ \ 
C \ = \ \left(
\begin{array}{llll}
 c_{11} &
   c_{12} &
   c_{13} &
   c_{14} \\
 c_{12} &
   -c_{11} &
   c_{14} &
   -c_{13} \\
 c_{31} &
   c_{32} &
   c_{33} &
   c_{34} \\
 c_{32} &
   -c_{31} &
   c_{34} &
   -c_{33}
\end{array}
\right) \eeq

We should require, see \eqref{eq:CR2}, that these are intertwined by any ${\bf \^K}$, i.e. by each of the $\^Y_\a$; again this can be written as 
$$ C \ = \ - \, \^Y_\a \, B \, \^Y_\a \ . $$
By explicit computations we conclude that 
\beq B \ = \ C \ = \ \left(
\begin{array}{llll}
 0 & 0 & k_3 & k_4
   \\
 0 & 0 & k_4 &
   - k_3 \\
 - k_3 & - k_4 & 0 &
   0 \\
 - k_4 & k_3 & 0 &
   0
\end{array}
\right) \ , \eeq
with $k_3 , k_4$ arbitrary constants. This relation can be rewritten as 
\beq B \ = \ C \ = \ k_3 \, Y_3 \ + \ k_4 \, Y_2 \ . \eeq

We can summarize our discussion as follows:

\medskip\noindent{\bf Lemma 1.} {\it The symmetries of the linear Dirac equation \eqref{eq:diracstandhh},  where $m \not= 0$ and $\pi$ should be considered as free parameters, are given by vector fields \eqref{eq:dirsymmM} with matrices $M$ of the block form
\beql{eq:Mam1} M \ = \ \left( \begin{array}{cc} 
k_0\ I + k_1 \^Y_2 & k_3 Y_3 + k_4 Y_2 \\ 
k_3 Y_3 + k_4 Y_2  & k_0\ I + k_1 \^Y_2 \end{array} \right)  \eeq
}

\medskip\noindent{\bf Remark 9.} This result confirms, of course, those given in the literature (see Theorem A above), albeit via a different approach.  If we write the symmetry matrices as in Section \ref{sec:symDirac1}, equation \eqref{ec:Diracmassiverep}, in the hyperhamiltonian approach we get the same result.\EOR


\subsection{Massless Dirac equation} 

In the massless case (Majorana-Weyl equation) the relations required by \eqref{eq:CR1} disappear, and we only have to deal with the conditions \eqref{eq:CR2} . Proceeding by direct computation, we obtain that 
\beql{eq:42AD} A \ = \ D \ = \ \left(
\begin{array}{llll}
 q_{4} & -q_{3} &
   q_{2} & -q_{1} \\
 q_{3} & q_{4} &
   -q_{1} & -q_{2} \\
 -q_{2} & q_{1} &
   q_{4} & -q_{3} \\
 q_{1} & q_{2} &
   q_{3} & q_{4}
\end{array}
\right) \eeq
\beql{eq:42BC} B \ = \ C \ = \ \left(
\begin{array}{llll}
 q_{8} & -q_{7} &
   q_{6} & -q_{5} \\
 q_{7} & q_{8} &
   -q_{5} & -q_{6} \\
 -q_{6} & q_{5} &
   q_{8} & -q_{7} \\
 q_{5} & q_{6} &
   q_{7} & q_{8}
\end{array}
\right) \ . \eeq

These are also written in the form 
\begin{eqnarray} A &=& D \ = \ q_4 \,I \ - \ q_3 \, Y_1 \ - \ q_1 \, Y_2 \ + \ q_2 \, Y_3 \ ,  \nonumber \\
B &=& C \ = \ q_8 \, Y \ - \ q_7 \, Y_1 \ - \ q_5 \, Y_2 \ + \ q_6 \, Y_3 \ . \label{eq:Mam2} \end{eqnarray}

\medskip\noindent
{\bf Remark 10.} The result is of course not surprising at all: we know that the two fundamental $\mathrm{su}(2)$ real irreducible representations are intertwining; see \cite{Kirillov}. \EOR
\bigskip

We can summarize our discussion as follows:

\medskip\noindent{\bf Lemma 2.} {\it The symmetries of the linear massless Dirac equation \eqref{eq:MW},  where $\pi$ should be considered as a free parameter, are given by vector fields \eqref{eq:dirsymmM} with matrices $M$ of the block form \eqref{eq:Mam0} and its block components given by \eqref{eq:Mam2}.}

\medskip\noindent{\bf Remark 11.} Again this result confirms, of course, those present in the literature (see Theorem B above), albeit via a different approach.  See also equation \eqref{ec:Diracmasslessrep}.\EOR

\section{Nonlinear Dirac equations and their symmetry} 

Finally, we note that nonlinear versions of the Dirac equation (describing self-interacting free fermions, with coupling constant $\kappa$) also exists and are quite relevant \cite{Thirring,Coleman,Ivanenko,Soler,NgPar,GrossNeveu}. We mention in particular what can be considered as the standard nonlinear Dirac equation 
\beql{eq:nldirac} \left(\ri \, \hbar \ \gamma^\mu \, \pa_\mu \ - \ m \, c \right) \ \psi \ = \ \kappa \, \left( \psi^+ \, \gamma^0 \, \gamma_\mu \, \gamma^5 \, \psi \, \gamma^\mu \, \gamma^5 \right) \ \psi \ , \eeq
originating in the Thirring model \cite{Thirring,Coleman}; but one also considers a different nonlinear Dirac equation, originating in the Ivanenko-Soler model \cite{Ivanenko,Soler}
\beql{eq:nldiracIS} \left(\ri \, \hbar \ \gamma^\mu \, \pa_\mu \ - \ m \, c \right) \ \psi \ = \ \kappa \, \left( \psi^+ \gamma^0 \, \psi \right) \ \psi \ . \eeq
In the one-dimensional case, this is also known as the Gross-Neveu equation \cite{GrossNeveu}.

These share the same symmetries of the linear Dirac equation, basically because the nonlinear terms transform in the same way as the linear ones, as required by their physical meaning; see also below for some information in this regard. Thus our results also extend to nonlinear Dirac equations.

It is maybe worth recalling how the nonlinear Dirac equations arise, in particular since, as mentioned above, one could and indeed does envisage different nonlinear extensions of the linear Dirac equation. 

We will quote from Fushchich and Zhdanov \cite{FZbook}

\medskip\noindent
{\it The first paper devoted to a nonlinear generalization of the Dirac equation was published by Ivanenko in 1938 \cite{FZ192}. Later Finkelstein with collaborators in 1951 \cite{FZ80,FZ81} and Heisenberg in 1953 \cite{FZ180,FZ181} started analyzing various nonlinear generalizations of the Dirac equation. Heisenberg \cite{FZ181,FZ182,FZ183,FZ184} put forward a vast program on the construction of a unified field theory of elementary particles. As a basis of this theory he chose a self-interacting spinor field described by a nonlinear equation. According to Heisenberg such a field is determined by} [eq.\eqref{eq:nldirac} above]. {\it We will call} [this] {\it system the Dirac-Heisenberg equation.}
\bigskip

In their book, Fushchich and Zhdanov reversed the usual approach, and rather than studying the symmetries of a given Nonlinear Dirac equation, they studied nonlinear extensions of the Dirac equations classifying them according to their symmetry properties.
Such a general extension of the linear Dirac equation is written in the form
\beql{eq:nldFZ}\ri \ \gamma^\mu \, \pa_\mu \, \psi \ = \ F (\psi^*,\psi ) \ . \eeq

The results about symmetries of equations in this general form for different choices of the function $F$ in \eqref{eq:nldFZ} are summarized in \cite{FZbook}, see in particular theorems 1.2.2 and 1.2.3 therein; proofs are given in \cite{FZ152,FZ155,FZbook}.

We will only mention that their discussion guarantees that  \eqref{eq:nldirac} and \eqref{eq:nldiracIS} are invariant under the same (Lie-point, local) symmetry algebra of the linear Dirac equation; this holds both for $m \not= 0$ and for the massless case $m=0$.


 \section{Conclusions}

The Dirac equation is one of the cornerstones of fundamental Physics; as such it and its properties have been thoroughly studied. In particular, for what concern its symmetries, it is well known that while it was built by Dirac primarily on the basis of relativistic invariance -- hence to be symmetric under the $P(1,3)$ Poincar\'e group -- in the massive case it also displays further ``special'' symmetries, described by the subgroup $V(4)$ given in Theorem A; in the massless case the ``built-in'' symmetry is not just the Poincar\'e group but the full Conformal group $C(1,3)$, but also in this case there are ``special'' symmetries, described by the subgroup $V(8)$ detailed in Theorem B (which includes the $V(4)$ subgroup). 

In previous works \cite{GM,MT,GR0,GR1,GR2,GR3,GR4,GR5} an extension of standard Hamiltonian mechanics based on hyperKahler structures, rather than on symplectic ones, was considered; this is called hyperhamiltonian (see e.g. \cite{GRbook} for a thorough review of this and of its applications). By construction, this is specially suited for studying problems related to spin structure, and in fact it turns out that -- among others -- the Dirac equation can be cast in hyperhamiltonian form \cite{GRDirac}. 

In this work we have discussed a question which is quite natural from this point of view, i.e. if the possibility of casting the \emph{symmetries} of the Dirac equation in hyperhamiltonian terms also exists. The question actually only arises for the $V(4)$ and $V(8)$ subgroups mentioned above: in fact, the dilation symmetry (related to the linear nature of the equation) can surely \emph{not} be described in hyperhamiltonian terms -- nor in hamiltonian ones -- as hyperhamiltonian vector fields preserve the volume { in the space in which they act (the space of wavefunctions in this case)}, while the Lorentz transformations can be described in quaternionic terms, as already { remarked and as already} known by Dirac \cite{DL}. We have thus tackled this question for both the $V(4)$ and $V(8)$ groups; in both cases we found that all vector fields which are \emph{not} dilations (thus, in our notation, at the exception of $Q_0$) can indeed be cast in hyperhamiltonian form, and provided an explicit form for their hyperhamiltonian representation. 

{ It should be noted that we have proved the hyperhamiltonian character of the Dirac equation using the momentum representation \cite{GRDirac}; this choice was needed since strictly speaking hyperhamiltonian dynamics is defined for ODEs, and the Dirac equation is in ODE form in the momentum representation and in PDE form in the coordinate representation. It is thus natural to wonder if our result also holds in the momentum representation. This was shown \emph{not} to be the case in Section \ref{sec:hhsymm}. Thus we have some kind of duality: in the coordinate representation it is possible to cast in hyperhamiltonian form the symmetries of the Dirac equation, but not the equation itself, while in the momentum representation we can cast in hyperhamiltonian form the Dirac equation, but not its symmetries.} 

Finally, we noted that nonlinear versions of the Dirac equation also exists and are quite relevant \cite{Thirring,Coleman,Ivanenko, Soler,NgPar,GrossNeveu}. These share the same symmetries of the linear Dirac equation, basically because their physical meaning requires that the nonlinear terms transform in the same way as the linear ones. Thus our results also extend to nonlinear Dirac equations. 

All this confirms the value of the hyperhamiltonian approach for studying spin-related problems, and in particular for the analysis of the (linear and nonlinear) Dirac equation.


\begin{appendix}


%
%
%


\section{A review of linear symmetries of the Dirac equation \label{sec:symDirac2}}


In this Appendix we will give detailed computations of the symmetries in the massive and massless cases and for different representations of Dirac $\gamma$ matrices.

\subsection{Massive case}

The Majorana and standard representations of the $\gamma$ matrices are used to find the symmetries of the Dirac equation in the massive case.

\subsubsection{Majorana representation}
Due to the well known fact that Dirac $\gamma$ matrices  are pure imaginary in the Majorana representation  ($M$), this is a convenient form to attack the problem, working with both the equations for $\psi$ and  its complex conjugate $\psi^*$. Consider the Dirac equation:
\beq
(\ri\,\hbar\,\gamma_M^{\mu}\,\pa_{\mu}-m\,c\,I_4)\,\psi=0
\eeq
and the Majorana representation (pure imaginary matrices):
\beq
\begin{aligned}
&\gamma_M^0=\begin{pmatrix}0 & \sigma^2\\ \sigma^2 & 0\end{pmatrix},\quad
\gamma_M^1=\begin{pmatrix}\ri\,\sigma^3 & 0\\ 0 &\ri\,\sigma^3\end{pmatrix},\\
&\gamma_M^2=\begin{pmatrix}0 & -\sigma^2 \\ \sigma^2& 0\end{pmatrix},\quad
\gamma_M^3=\begin{pmatrix}-\ri\, \sigma^1 & 0\\ 0 & -\ri\,\sigma^1  \end{pmatrix}.
\end{aligned}
\eeq

Construct the complex conjugate of the equation \eqref{eq:dirac}, taking into account that $(\gamma_M^{\mu})^*=-\gamma_M^{\mu}$:
\beq
(\ri\,\hbar\,\gamma_M^{\mu}\,\pa_{\mu}-m\,c\,I_4)\,\psi^*=0\ ,
\eeq
which is (obviously, since the equation is real except for the wave function $\psi$)  the same equation.
Make a linear transformation of the wave function and  its complex conjugate (which is another way to face the problem of constructing a real Dirac equation):
\beq
\psi'=A\psi+B\psi^*
\eeq
and then
\beq
\psi'^*=A^*\psi^*+B^*\psi
\eeq
or, in matrix form:
\beq
\begin{pmatrix} \psi'\\ \psi'^*\end{pmatrix}=
\Lambda
\begin{pmatrix} \psi\\ \psi^*\end{pmatrix},\quad \Lambda=\begin{pmatrix} A & B \\ B^* & A^*\end{pmatrix}\ .
\eeq

When $m\neq 0$, the equations are::
\beq
\begin{pmatrix}\gamma_M^{\mu}\pa_{\mu} & 0 \\ 0 & \gamma_M^{\mu}\pa_{\mu} \end{pmatrix}\begin{pmatrix} \psi\\ \psi^*\end{pmatrix}=\frac{m\,c}{\ri\,\hbar} \begin{pmatrix} \psi\\ \psi^*\end{pmatrix}\ ,
\eeq
and, in the prime coordinates:
\beq
\begin{pmatrix}\gamma_M^{\mu}  & 0 \\ 0 & \gamma_M^{\mu}  \end{pmatrix}\Lambda^{-1}\pa_{\mu}\begin{pmatrix} \psi'\\ \psi'^*\end{pmatrix}=\frac{m\,c}{\ri\,\hbar} \Lambda^{-1}\begin{pmatrix} \psi'\\ \psi'^*\end{pmatrix}
\eeq
or
\beq
\Lambda\begin{pmatrix}\gamma_M^{\mu}  & 0 \\ 0 & \gamma_M^{\mu}  \end{pmatrix}\Lambda^{-1}\pa_{\mu}\begin{pmatrix} \psi'\\ \psi'^*\end{pmatrix}=\frac{m\,c}{\ri\,\hbar} \begin{pmatrix} \psi'\\ \psi'^*\end{pmatrix}\ .
\eeq

Imposing invariance:
\beql{eq:invp}
\Lambda\begin{pmatrix}\gamma_M^{\mu}  & 0 \\ 0 & \gamma_M^{\mu}  \end{pmatrix}\Lambda^{-1}=\begin{pmatrix}\gamma_M^{\mu}  & 0 \\ 0 & \gamma_M^{\mu}  \end{pmatrix}
\eeq
or, using $(\gamma_M^{\mu})^*=-\gamma_M^{\mu}$:
\beq
[\gamma_M^{\mu},A]=[\gamma_M^{\mu},B]=0,\quad \mu=0,1,2,3\ .
\eeq

Then, since the matrices $\gamma_M^{\mu}$ generate an irreducible representation of a Clifford algebra, by Schur's Lemma, 
\beq
A=\lambda\, I_4,\quad B=\mu\, I_4,\quad \lambda,\mu\in\mathbb{C}
\eeq
and the matrix $\Lambda$ is:
\beq
\Lambda=\begin{pmatrix} \lambda\, I_4 & \mu\, I_4 \\ \mu^*\, I_4 & \lambda^*\, I_4\end{pmatrix}.
\eeq
Taking  real parameters we find the following four linearly independent matrices ($I_4$ is the identity matrix in dimension $4$, the factor 1/2 simplifies the commutation relations of these matrices)\footnote{The matrices computed above can be considered as a basis of the Lie algebra of the transformation group.}:
\beql{eq:oneMaj}
I_8=\frac12\sigma^0\otimes I_4,\quad
J_1^M=\frac{\ri}{2}\,\sigma^3\otimes I_4,\quad
K_2^M=\frac12\sigma^2\otimes I_4,\quad
K_3^M=\frac12\sigma^1\otimes I_4,\quad
\eeq
 which provide four real vector fields in $\psi,\psi^*$ (written in the Majorana representation):
\beq
\begin{aligned}
X_0^M &= \frac12\left(\psi\,\pa_{\psi}+\psi^*\,\pa_{\psi^*}\right),\\
X_1^M &= \frac{\ri}{2}\,\left(\psi\,\pa_{\psi}-\psi^*\,\pa_{\psi^*}\right),\\
X_2^M &= -\frac{\ri}{2}\,\left(\psi^*\,\pa_{\psi }-\psi\,\pa_{\psi^*}\right),\\
X_3^M &= \frac12\left(\psi^*\,\pa_{\psi}+\psi\,\pa_{\psi^*}\right).
\end{aligned}
\eeq

The commutation table of the four matrices \eqref{eq:oneMaj} is presented  in Table \ref{table:tab3}:
\begin{table}
\begin{center}
\begin{tabular}{c|cccc}
& $I_8$& $J_1$& $K_2$& $K_3$\\
\midrule
$I_8$ & $0$ & $0$    & $0$   & $0$\\
$J_1$ & $0$ & $0$    & $K_3$ & $-K_2$\\
$K_2$ & $0$ & $-K_3$ & $0$   & $-J_1$\\
$K_3$ & $0$ & $K_2$  & $J_1$ & $0$ 
\end{tabular}
\caption{Commutation table of the matrices of the symmetry algebra  of the massive Dirac equation.\label{table:tab3}}
\end{center}\end{table}
\noindent showing that they form  a basis of a real Lie algebra of dimension 4, which is in fact isomorphic to $\mathfrak{g}\mathfrak{l}(2,\mathbb{R})$ (or, the direct sum of a one-dimensional center plus the Lie algebra of the Lorentz group in $1+2$ dimensions). The notation will be justified when we consider the massless case below.

\subsubsection{Standard representation}

The above vector fields are not those given in \cite{FZbook}. However, all the representations of Dirac $\gamma$ matrices are unitarily equivalent and the explicit relation between the Dirac ($D$) and Majorana representations is
\beq
\gamma_D^{\mu}=U\gamma_M^{\mu}U^+,\quad \mu=0,1,2,3,\quad U=\frac{1}{\sqrt{2}}\begin{pmatrix}1 & 0 & 0 & -\ri \\ 0 & 1 & \ri & 0 \\ 0 & -\ri & -1 & 0 \\ \ri &0 & 0 & -1  \end{pmatrix}\ .
\eeq
where
\beq
\gamma_D^0=\begin{pmatrix}\sigma^0 & 0\\ 0 & -\sigma^0\end{pmatrix},\;
\gamma_D^1=\begin{pmatrix}0 & \sigma^1\\  -\sigma^1 &0\end{pmatrix},\;
\gamma_D^2=\begin{pmatrix}0 & \sigma^2 \\ -\sigma^2& 0\end{pmatrix},\;
\gamma_D^3=\begin{pmatrix}0 &  \sigma^3\\  -\sigma^3 & 0  \end{pmatrix}.
\eeq

Then
\beq
\begin{pmatrix}\psi_M \\ \psi_M^*\end{pmatrix}=\tilde{U}
\begin{pmatrix}\psi_D \\ \psi_D^*\end{pmatrix},\quad \tilde{U}=\begin{pmatrix} U & 0  \\ 0  & U^*\end{pmatrix}\ .
\eeq

The matrices $L_{i}^M$, $i=1,2,3,4$, in the Majorana representation are transformed into ($U^+=U$):
\beq
\tilde{U}\Lambda\tilde{U}^+=\begin{pmatrix} U & 0  \\ 0  & U^*\end{pmatrix}\begin{pmatrix} \lambda\, I_4 & \mu\, I_4 \\ \mu^*\, I_4 & \lambda^*\, I_4\end{pmatrix}\begin{pmatrix} U & 0  \\ 0  & U^*\end{pmatrix}
=\begin{pmatrix} \lambda\, I_4 & \mu\, UU^* \\ \mu^*\, U^*U & \lambda^*\, I_4\end{pmatrix} ,
\eeq
where
\beq
UU^*=-U^*U=\begin{pmatrix} 0 & -\sigma^2 \\ \sigma^2 & 0\end{pmatrix}=-\gamma_D^2\ ,
\eeq
and then,
\beq
\tilde{U}\Lambda\tilde{U}^+=\begin{pmatrix} \lambda\, I_4 & -\mu\, \gamma_D^2 \\ \mu^*\, \gamma_D^2 & \lambda^*\, I_4\end{pmatrix}\ .
\eeq
In terms of real parameters we have the four matrices (with, obviously, the same commutation rules as those in Table \ref{table:tab4}):
\beq
I_8=\frac12\sigma^0\otimes I_4,\quad
J_1^D= \frac{\ri}{2}\,\sigma^3\otimes I_4, \quad
K_2^D= \frac{\ri}{2}\, \sigma^1\otimes \gamma_D^2 ,\quad
K_3^D=-\frac{\ri}{2}\,\sigma^2\otimes \gamma_D^2, 
\eeq
and the real vector fields (where the wave functions are written in the standard representation)
\beq
\begin{aligned}
X^D_0&=\frac12(\psi\,\pa_{\psi}+\psi^*\,\pa_{\psi^*}),\\
X^D_1&=\frac{\ri}{2}\left(\psi\,\pa_{\psi}-\psi^*\,\pa_{\psi^*}\right),\\
X^D_2&= \frac{\ri}{2}\,\left((\gamma^2_D\psi^*)\,\pa_{\psi}+(\gamma^2_D\psi )\,\pa_{\psi^*}\right) ,\\
X^D_3&=\frac{1}{2}\,\left(-(\gamma^2_D\psi^*)\,\pa_{\psi}+(\gamma^2_D\psi)\,\pa_{\psi^*}\right),
\end{aligned}
\eeq
which are the vector fields in \cite{FZbook} (up to order and global factors).


\subsection{Massless case}

We now turn to the massless case, again in the Majorana and standard representations.

\subsubsection{Majorana representation}
Consider the massless Dirac equation ($m=0$):
\beq
\gamma_M^{\mu}\,\pa_{\mu}\,\psi=0 \ ,
\eeq
again with Majorana matrices.

Construct the complex conjugate of the equation:
\beq
\gamma_M^{\mu}\,\pa_{\mu}\,\psi^*=0\ , 
\eeq
which is, as in the massive case, the same equation for the wave function $\psi^*$. Under a linear transformation:
\beq
\psi'=A\psi+B\psi^*,\quad \psi'^*=A^*\psi^*+B^*\psi\ ,
\eeq
or, in matrix form:
\beq
\begin{pmatrix} \psi'\\ \psi'^*\end{pmatrix}=
\Lambda
\begin{pmatrix} \psi\\ \psi^*\end{pmatrix},\quad \Lambda=\begin{pmatrix} A & B \\ B^* & A^*\end{pmatrix}\ ,
\eeq
the equations are transformed as:
\beq
\begin{pmatrix}\gamma_M^{\mu} & 0 \\ 0 & \gamma_M^{\mu} \end{pmatrix}\pa_{\mu}\begin{pmatrix} \psi\\ \psi^*\end{pmatrix}=0
,\quad 
\begin{pmatrix}\gamma_M^{\mu}  & 0 \\ 0 & \gamma_M^{\mu}  \end{pmatrix}\Lambda^{-1}\pa_{\mu}\begin{pmatrix} \psi'\\ \psi'^*\end{pmatrix}=0\ .
\eeq

Imposing invariance:
\beq
\begin{pmatrix}\gamma_M^{\mu}  & 0 \\ 0 & \gamma_M^{\mu}  \end{pmatrix}\Lambda^{-1}=K \begin{pmatrix}\gamma^{\mu}_M  & 0 \\ 0 & \gamma^{\mu}_M  \end{pmatrix},\
\eeq
where $K$ is a regular matrix. Let us write:
\beq
K^{-1}=\begin{pmatrix}C_1  & C_2 \\ C_3 & C_4  \end{pmatrix}\ .
\eeq 
Then, for $\mu=0,1,2,3$:
\beq
\begin{pmatrix} A & B \\ B^* & A^*\end{pmatrix}=\begin{pmatrix}(\gamma^{\mu}_M)^{-1}\,C_1\,\gamma_M^{\mu}  & (\gamma^{\mu}_M)^{-1}\,C_2\,\gamma^{\mu}_M \\ (\gamma^{\mu}_M)^{-1}\,C_3\,\gamma^{\mu}_M & (\gamma^{\mu}_M)^{-1}\,C_4\,\gamma^{\mu}_M  \end{pmatrix}\ .
\eeq
Since this relation should be true for $\mu=0,1,2,3$, we get:
\beq
(\gamma^0_M)^{-1}\,C_1\gamma^{\mu}_M=(\gamma^1_M)^{-1}\,C_1\gamma^{1}_M=(\gamma^2_M)^{-1}\,C_1\gamma^{2}_M=(\gamma^3_M)^{-1}\,C_1\gamma^{3}_M\ ,
\eeq
or, using the properties of the $\gamma$ matrices:
\beq
 \gamma^0_M \,C_1\gamma^{\mu}_M=-\gamma^1_M \,C_1\gamma^{1}_M=-\gamma^2_M \,C_1\gamma^{2}_M=-\gamma^3_M\,C_1\gamma^{3}_M\ .
\eeq
We get the same equation for $C_2$, and then, $C_3=C_2^*$, $C_4=C_1^*$. After a straightforward computation we obtain:
\beq
C_1=\lambda_1 I_4+\lambda_2\gamma^5_M,\quad
C_2=\mu_1 I_4+\mu_2\gamma^5_M,\quad \lambda_i,\mu_i\in\mathbb{C},\quad i=1,2\ .
\eeq
The matrix $\Lambda$ is:
\beq
\Lambda=\begin{pmatrix} \lambda_1\,I_4+\lambda_2\gamma^5_M & \mu_1\, I_4+\mu_2\gamma^5_M \\[10pt] \mu_1^*\,I_4-\mu_2^*\,\gamma^5_M & \lambda_1^*\,I_4-\lambda_2^*\,\gamma^5_M\end{pmatrix}\ .
\eeq
Considering real parameters we get the eight matrices:
\beql{ec:mat1}
I_8=\frac12\,\sigma^0\otimes I_4,\quad
J_1^M=\frac{\ri}{2}\,\sigma^3\otimes I_4,\quad
K_2^M=\frac12\,\sigma^2\otimes I_4,\quad
K_3^M=\frac12\,\sigma^1\otimes I_4, 
\eeq
\beql{ec:mat2}
L^M_8=\ri\,\sigma^0\otimes \gamma^5_M,\quad
J_2^M=\ri\, \sigma^2\otimes \gamma^5_M,\quad
J_3^M=\ri\, \sigma^1\otimes \gamma^5_M,\quad
K_1^M=\frac12\sigma^3\otimes \gamma^5_M,
\eeq
providing the following eight real vector fields in $\psi,\psi^*$:
\beq
\begin{aligned}
X_0^M &= \frac12\left(\psi\,\pa_{\psi}+\psi^*\,\pa_{\psi^*}\right),\\
X_1^M &= \frac{\ri}{2}\,\left(\psi\,\pa_{\psi}-\psi^*\,\pa_{\psi^*}\right),\\
X_2^M &= -\frac{\ri}{2}\,\left(\psi^*\,\pa_{\psi }-\psi\,\pa_{\psi^*}\right),\\
X_3^M &= \frac12\left(\psi^*\,\pa_{\psi}+\psi\,\pa_{\psi^*}\right),\\
Z_0^M &= \frac{\ri}{2}\,\left((\gamma^5_M \psi)\,\pa_{\psi}+(\gamma^5_M \psi^*)\,\pa_{\psi^*}\right),\\
Z_1^M &=  \frac{1}{2}\left((\gamma^5_M \psi^*)\,\pa_{\psi}-(\gamma^5_M \psi)\,\pa_{\psi^*} \right),\\
Z_2^M &=  \frac{\ri}{2}\left((\gamma^5_M \psi^*)\,\pa_{\psi}+(\gamma^5_M \psi)\,\pa_{\psi^*}\right),\\
Z_3^M &= \frac{1}{2}\,\left((\gamma^5_M \psi)\,\pa_{\psi}-(\gamma^5_M \psi^*)\,\pa_{\psi^*}\right).
\end{aligned}
\eeq
The vector fields $X_0^M,\ldots ,X_3^M$ are those obtained in the massive case. 

The Lie algebra  generated by the matrices \eqref{ec:mat1} and \eqref{ec:mat2} have commutation rules given in  Table \ref{table:tab2}. The Lie algebra is isomorphic to the Lie algebra of the Lorentz group plus a center of dimension 2.

\begin{table}
\begin{center}
\begin{tabular}{c|cccccccc}
& $J_1$& $J_2$& $J_3$&  $K_1$& $K_2$& $K_3$\\
\midrule
$J_1$ & $0$    & $J_3$   & $-J_2$ & $0$    & $K_3$  & $-K_2$ \\
$J_2$ & $-J_3$ & $0$     & $J_1$  & $-K_3$ & $0$    & $K_1$  \\
$J_3$ & $J_2$  & $-J_1$  & $0$    & $K_2$  & $-K_1$ & $0$    \\
$K_1$ & $0$    & $K_3$   & $-K_2$ & $0$    & $-J_3$ & $J_2$  \\
$K_2$ & $-K_3$ & $0$     & $K_1$  & $J_3$  & $0$    & $-J_1$ \\
$K_3$ & $K_2$  & $-K_1$  & $0$    & $-J_2$ & $J_1$  & $0$
\end{tabular}
\caption{Commutation table of the matrices of symmetries for the massless Dirac equation.\label{table:tab4}}
\end{center}\end{table}

\subsubsection{Standard representation}

As in the massive case, the symmetry vector fields are not those given in \cite{FZbook}. Using the matrix $\tilde{U}$,  
the matrices $L_{i}^M$, $i=1,\ldots,8$, in the Majorana representation are transformed into:
\begin{eqnarray*}
\tilde{U}\,\Lambda\,\tilde{U}^+&=&\begin{pmatrix} U & 0  \\ 0  & U^*\end{pmatrix}
\begin{pmatrix} \lambda_1\,I_4+\lambda_2\,\gamma^5_M & \mu_1\,I_4+\mu_2\,\gamma^5_M \\[10pt] \mu_1^*\,I_4-\mu_2^*\,\gamma^5_M & \lambda_1^*\,I_4-\lambda_2^*\,\gamma^5_M\end{pmatrix}
\begin{pmatrix} U & 0  \\ 0  & U^*\end{pmatrix}^{+} \\
 &=&
\begin{pmatrix} \lambda_1\,I_4+\lambda_2\,U\,\gamma_M^5\,U & \mu_1\,U\,U^*+\mu_2\,U\,\gamma^5_M\,U^* \\[10pt] \mu_1^*\,U^*\,U-\mu_2^*\,U^*\,\gamma_M^5\,U & \lambda_1^*\,I_4-\lambda_2^*\,U^*\,\gamma^5_M\,U^*\end{pmatrix}
\end{eqnarray*}
Simplifying
\beq 
U\,\gamma_M^5\,U=\gamma_D^5,\quad U^*\,\gamma^5_M\,U^*=-\gamma^5_D,\quad U\,U^*=-\gamma_D^2,\quad U^*\,U=\gamma_D^2
\eeq
$$ 
U\,\gamma_M^5\,U^*=U^*\,\gamma^5_M\,U=\gamma^2_D\gamma^5_D
$$
and then,
$$
\tilde{U}\,\Lambda\,\tilde{U}^+
=
\begin{pmatrix} \lambda_1\,I_4+\lambda_2\, \gamma_D^5  & -\mu_1\,\gamma_D^2+\mu_2\,\gamma_D^2\,\gamma^5_D \\[10pt] \mu_1^*\,\gamma_D^2-\mu_2^*\,\gamma_D^2\,\gamma_D^5 & \lambda_1^*\,I_4+\lambda_2^*\, \gamma^5_D \end{pmatrix}
$$
In terms of real parameters we have the eight matrices:
\beq
\begin{aligned}
&I_8=\frac12\sigma^0\otimes I_4,&&
J_1^D= \frac{\ri}{2}\,\sigma^3\otimes I_4, \\
&K_2^D= \frac{\ri}{2}\, \sigma^1\otimes \gamma_D^2 ,&&
K_3^D=-\frac{\ri}{2}\,\sigma^2\otimes \gamma_D^2, 
\\
&L^D_8=\frac{\ri}{2}\sigma^3\otimes \gamma_D^5,&&
J^D_2=\frac{\ri}{2}\sigma^2\otimes (\gamma_D^2\gamma_D^5),\\
&J^D_3=\frac{\ri}{2}\sigma^1\otimes (\gamma_D^2\gamma_D^5),&&
K^D_1=\frac{1}{2}\sigma^0\otimes \gamma^5_D,\quad
\end{aligned}
\eeq
which correspond to the eight real vector fields given in \cite{FZbook} (up to order and global factors):
\beq
\begin{aligned}
X^D_0&=\frac12(\psi\,\pa_{\psi}+\psi^*\,\pa_{\psi^*}),\\
X^D_1&=\frac{\ri}{2}\left(\psi\,\pa_{\psi}-\psi^*\,\pa_{\psi^*}\right),\\
X^D_2&= \frac{\ri}{2}\,\left((\gamma^2_D\psi^*)\,\pa_{\psi}+(\gamma^2_D\psi )\,\pa_{\psi^*}\right) ,\\
X^D_3&=\frac{1}{2}\,\left(-(\gamma^2_D\psi^*)\,\pa_{\psi}+(\gamma^2_D\psi)\,\pa_{\psi^*}\right),\\
Z_0^D &= \frac{\ri}{2}\,\left((\gamma^5_D \psi)\,\pa_{\psi}-(\gamma^5_D \psi^*)\,\pa_{\psi^*}\right),\\
Z_1^D &= \frac{1}{2}\,\left((\gamma^2_D\gamma^5_D \psi^*)\,\pa_{\psi}-(\gamma^2_D\gamma^5_D \psi)\,\pa_{\psi^*}\right),\\
Z_2^D &= \frac{\ri}{2}\,\left((\gamma^2_D\gamma^5_D \psi^*)\,\pa_{\psi}+(\gamma^2_D\gamma^5_D \psi)\,\pa_{\psi^*}\right) ,\\
Z_3^D &= \frac{1}{2}\left( ( \gamma^5_D \psi )\,\pa_{\psi}+( \gamma^5_D \psi^*)\,\pa_{\psi^*}\right).
\end{aligned}
\eeq


\section{Some observations about Fourier transform and symmetries\label{sec:Fourier}}

In order to justify the results corresponding to the partial Fourier transform in the Dirac equation \eqref{eq:diracstandhh} we detail in this Appendix the necessary computations.

Consider a linear first order partial differential equation (such as the Dirac equation) written as an evolution equation in the momentum space:
\beq
\pa_t\psi=\sum_{j=1}^3L_j\pa_j\psi+m L_0\psi
\eeq
where $L_{\mu}$, $\mu=0,1,2,3$ are some constant matrices and $\psi$ a vector valued function. As we have discussed, a linear transformation:
\beq
\psi'=\Lambda\psi
\eeq
is a symmetry of the equation if
\beq
[\Lambda,L_{\mu}]=0,\quad \mu=0,\ldots 3
\eeq
 
Take the Fourier transform of $\psi$ in the spatial coordinates:
\beq
\psi(t,\mathbf{x})=\frac{1}{(2\pi)^{3/2}}\int_{\mathbb{R}^3}\wdh{\psi}(t,\mathbf{p})\,\mathrm{e}^{\ri\, \mathbf{p}\cdot\mathbf{x}}\,\mathrm{d}\mathbf{x}
\eeq
Then, substituting in the equation: 
\beq
\pa_t \wdh{\psi} =\ri\sum_{j=1}^3L_j  p_j\wdh{\psi} +m L_0 \wdh{\psi} 
\eeq
Since Fourier transform is a linear transformation the symmetries (that is, linear transformations of the wave function or its Fourier transform) are the same.

The complex conjugate of this equation is: 
\beq
\pa_t \wdh{\psi}^* =-\ri\sum_{j=1}^3 L_j^*  p_j\wdh{\psi}^* +m L_0^* \wdh{\psi}^* 
\eeq
and we can construct the system:
\beq
\pa_t\begin{pmatrix}\wdh{\psi}\\ \wdh{\psi}^*\end{pmatrix}=
\begin{pmatrix}\ri\sum_{j=1}^3L_j  p_j+m L_0 & 0\\ 0  & -\ri\sum_{j=1}^3L_j^*  p_j+m L_0^*\end{pmatrix}\begin{pmatrix}\wdh{\psi}\\ \wdh{\psi}^*\end{pmatrix}
\eeq
or,
\beq
\pa_t\begin{pmatrix}\wdh{\psi}\\ \wdh{\psi}^*\end{pmatrix}=
H\begin{pmatrix}\wdh{\psi}\\ \wdh{\psi}^*\end{pmatrix},\quad H=\ri\sum_{j=1}^3p_j\begin{pmatrix}L_j    & 0\\ 0  & -L_j^*   \end{pmatrix}+m\begin{pmatrix}L_0 & 0\\ 0  & L_0^*\end{pmatrix}
\eeq

Consider a linear transformation mixing $\psi$ and $\psi^*$:
\beq
\begin{pmatrix}\wdh{\psi}\\ \wdh{\psi}^*\end{pmatrix}=\Lambda
\begin{pmatrix}\wdh{\psi}\\ \wdh{\psi}^*\end{pmatrix},\quad \Lambda=\begin{pmatrix}A & B\\ B^*  & A^*\end{pmatrix}
\eeq
In order that $\Lambda$ be a symmetry, it must commute with $H$, that is:
\beq
[H,\Lambda]=0,
\eeq
\beq
 \left[\begin{pmatrix}L_j    & 0\\ 0  & -L_j^*   \end{pmatrix},\begin{pmatrix}A & B\\ B^*  & A^*\end{pmatrix}  \right]=0,\; j=1,2,3,\quad \left[ \begin{pmatrix}L_0   & 0\\ 0  & L_0^*   \end{pmatrix},\begin{pmatrix}A & B\\ B^*  & A^*\end{pmatrix}\right]=0
\eeq 
 In an explicit form:
\beq
[L_j,A]=0,\quad       L_jB=-B L_j^*,\quad   j=1,2,3,
\eeq
\beq
 [L_0,A]=0 ,\quad  L_0B=BL_0^*  
\eeq 

In the Dirac equation the matrices $L_{\mu}$ are $L_0=\gamma^0$, $L_i=(\gamma^0)^{-1}\gamma^i=\gamma^0 \gamma^i$. Then:
\beq
[ \gamma^{0}  \gamma^j,A]=0,\quad  [\gamma^0,A]=0. 
\eeq 
and, since the matrices $\gamma^{\mu}$ generate an irreducible representation of a Clifford algebra,
\beq
A=\lambda I_4,\quad \lambda \in \mathbb{C}
\eeq
In the Majorana representation,
\beq
L_0^*=-L_0,\quad L_j^*=L_j,\quad j=1,2,3
\eeq
and $B$ satisfies the equations:
\beq
  L_jB+B L_j=0,\quad   j=1,2,3,\quad 
  [L_0,B]=0  
\eeq 
with solution:
\beq
B=\mu\,\gamma_M^0,\quad \mu\in   \mathbb{C}
\eeq
Finally, the matrix $\Lambda $ is:
\beq
\Lambda=\begin{pmatrix}\lambda\, I_4 & \mu\,\gamma_M^0\\ -\mu^*\,\gamma_M^0  & \lambda^*\,I_4\end{pmatrix}
\eeq
that is, the four independent matrices:
\beq
\frac12\,\begin{pmatrix}  I_4 & 0\\ 0  & I_4\end{pmatrix},\quad
\frac{\ri}{2}\,\,\begin{pmatrix} I_4 & 0 \\ 0  & -I_4\end{pmatrix},\quad 
\frac12\,\begin{pmatrix} 0 & \gamma_M^0\\  -\gamma_M^0  & 0\end{pmatrix},\quad 
\frac{\ri}{2}\,\begin{pmatrix} 0 & \gamma_M^0\\ \gamma_M^0  & 0 \end{pmatrix}, 
\eeq
or
\beq 
\wdh{L}_0^M=\frac12\sigma^0\otimes I_4,\quad
\wdh{L}_1^M=\frac{\ri}{2}\,\sigma^3\otimes I_4,\quad
\wdh{L}_2^M=\frac{1}{2}\sigma^2\otimes (\ri\,\gamma_M^0),\quad
\wdh{L}_3^M=\frac{1}{2}\sigma^1\otimes (\ri\,\gamma_M^0), 
\eeq

If we compare these matrices with the matrices we got for the Dirac equation in the configuration space \eqref{eq:oneMaj}:
\beq 
I_8=\frac12\sigma^0\otimes I_4,\quad
J_1^M=\frac{\ri}{2}\,\sigma^3\otimes I_4,\quad
K_2^M=\frac12\sigma^2\otimes I_4,\quad
K_3^M=\frac12\sigma^1\otimes I_4,\quad
\eeq
we observe that for two of these matrices (those corresponding to boosts) the identity matrix has been changed to $\ri\,\gamma_M^0$ which is a consequence of the Fourier transformation in the spatial variables.

\end{appendix}

\end{document}